# Calibration of quantum detector of noise based on a system of asymmetric superconducting loops

V.L. Gurtovoi, S.V. Dubonos, A.V. Nikulov, N.N. Osipov, and V.A. Tulin
(gurtovoi@ipmt-hpm.ac.ru, nikulov@ipmt-hpm.ac.ru )
Institute of Microelectronics Technology, Russian Academy of Sciences
142432 Chernogolovka, Moscow District, Russia

**ABSTRACT**

The quantum oscillations $V(\Phi/\Phi_0)$ of the dc voltage are induced on segments of asymmetric superconducting loops by an external ac current or noise. The dependencies of the amplitude of $V(\Phi/\Phi_0)$ on amplitude of inducing ac current are measured at different temperatures below superconducting transition $T_c$ on aluminum asymmetric loops and systems of the loops connected in series. The measured values of the maximum amplitude of the quantum oscillations $V(\Phi/\Phi_0)$, the amplitude of the ac current inducing this maximum dc voltage and the critical amplitude of the ac current decrease with temperature increase to $T_c$. The extrapolation of these measured dependencies to the region near superconducting transition allows to make a calibration of asymmetric superconducting loops as quantum detector of noise. The calibration restores an amplitude profile of the noise pulses from a measured temperature dependence of an amplitude of the quantum oscillations $V(\Phi/\Phi_0)$ induced by this noise. It is found that rectification efficiency, determined as relation of the maximum amplitude of the quantum oscillations $V(\Phi/\Phi_0)$ to the ac current amplitude inducing it, decreases near superconducting transition $T_c$. High efficiency of rectification observed below $T_c$ is consequence of irreversibility of the current-voltage curves. Increase of the rectification efficiency is achieved in multiple series connected loop structures.

## 1. INTRODUCTION

The advancement of technologies allows to realize precision devices very sensitive to electronic noise and which can not effectively work without precision control of external and internal noises. One of the most grandiose ideas of the last time is the idea of quantum computations[1]. The future realization of quantum computer is closely connected with solid-state quantum bit (qubit) and coupled qubit fabrication and investigation of their functionality. The most promising candidates for solid-state qubits are two-level superconducting structures which act like artificial atoms with quantized energy levels.[2-3] To carry out quantum computations, decoherence time for superposition of macroscopic quantum states in superconducting structures should be at least $10^4$ times longer than computation time.[4] Decoherence is determined by the equilibrium and non-equilibrium noise environment, which have to be investigated and strictly controlled. Although, the intensity of equilibrium thermal fluctuations vanishes linearly with temperature, nevertheless, there are quantum fluctuations due to zero-point motion even at zero temperature. To detect non-equilibrium and vanishing with temperature equilibrium thermal fluctuations and separate them from quantum fluctuations, one should use a *quantum system*, i.e. a quantum detector of noise (QDN).

As a basic element of the QDN an asymmetric superconducting loop (ASL) with or without Josephson junctions may be proposed to use. It was found that the dc potential difference $V(\Phi/\Phi_0)$ can be observed on segment of asymmetric superconducting loops with[5] and without[6,7] Josephson junctions. Sing and value of this dc voltage are periodical function $V(\Phi/\Phi_0)$ of magnetic flux $\Phi$ with period corresponding the flux quantum $\Phi_0 = \pi\hbar/e$ inside the loop[5-7]. This quantum oscillations can be observed without an evident power source in a temperature region near superconducting transition. At a lower temperature they can be induced by an external ac current, for example $I_{ext}(t) = I_0 \sin(2\pi ft)$, when its amplitude $I_0$ exceeds a critical value $I_{0,c}$[7]. It have been found that the dependence of the amplitude $V_A$ of the quantum oscillations on the amplitude $I_0$ of the external ac current is not monotonous. The amplitude of the voltage oscillations mounts quickly a maximum value $V_{max}$ at $I_0 = I_{max}$ and decrease further with increase of the amplitude of the ac current[7].

The external ac current with different frequency f in a wide spectrum was used and it have been found that the result does not depend on the frequency[7].

There is important that the critical amplitude $I_{0,c}$ of the ac current which can induce the dc voltage is close the superconducting critical current $I_c(T)$ of the loop. One can expect also that the maximum amplitude $V_{max}$ of the quantum oscillations of the dc voltage is proportional to the amplitude of the oscillations of the persistent current $I_p(\Phi/\Phi_0)$. The observed magnetic dependence of the dc voltage $V(\Phi/\Phi_0)$ [6,7] is like the one $I_p(\Phi/\Phi_0)$ of average equilibrium value of the persistent current. We can write, by analogy with the potential difference $V = (R_{ls} – R_l l_s/l)I = R_{asym}I$ observed of segment $l_s$, with a resistance $R_{ls}$ of conventional loop l with a resistance $R_l$ and a circular current I induced in a loop by the Faraday's voltage $R_l I = -d\Phi/dt$, that the potential difference is observed because of resistance difference $R_{asym} \neq 0$ of loop segments $V(\Phi/\Phi_0) = R_{asym}I_p(\Phi/\Phi_0)$. It is known that the critical current and the persistent current decrease down to zero near the critical temperature $T_c$. The both currents becomes zero at the critical temperature $T = T_c$ if thermal fluctuations are not taken into account. The thermal fluctuations decreases the critical current and increases the persistent current. It is known from theory[9] and experimental results[10] that because of thermal fluctuations the persistent current is observed not only below but also above the critical temperature. There is a fluctuation region near $T_c$ where the critical current equals zero $I_c = 0$ (i.e. an equilibrium resistance is not zero $R_l > 0$) and the persistent current is not zero $I_p \neq 0$. The first experimental evidence of $I_p \neq 0$ at $R_l > 0$ is the Little-Parks oscillations $R(\Phi/\Phi_0)$ of the resistance R of superconducting cylinder[10] or loop[11], which are observed only in a narrow fluctuation region near $T_c$. Therefore one can expect that any how weak noise, right down to the equilibrium noise[12], can induce the quantum oscillations of the dc voltage near critical temperature if the critical amplitude $I_{0,c}$ is close to the critical current and the maximum amplitude $V_{max}$ is proportional to the persistent current near $T_c$.

The equilibrium electric noise was measured first by J.B.Johnson[13] and described theoretically by H.Nyquist[14] and is known as Johnson[15] or Nyquist noise. There is important difference between power of the Nyquist's noise and the dc power: the first can not be summed and the second can. The power of the Nyquist's noise of N elements equals the one of a single element whereas the power of a system of N dc power source connected in series in N time higher than the one of the like single source. Therefore the transformation of the power of random noise into the dc power observed on asymmetric superconducting loops gives very important advantage: even very weak noise can give enough high output power. It was shown[7] that the dc voltage observed on asymmetric superconducting loops is summed in a system of the loops connected in series: the amplitude of the quantum oscillation $V(\Phi/\Phi_0)$ induced by an external ac current on a system of N asymmetric loops approximately in N time higher than the one observed on single loop. The summation of the dc voltage $V(\Phi/\Phi_0)$ and dc power was observed in[7] below $T_c$, in superconducting state, where the quantum oscillations of the dc voltage are induced be an external ac current. The same summation may be expected near superconducting transition, where $V(\Phi/\Phi_0)$ can be induced by an equilibrium and non-equilibrium noise. Therefore it was proposed[16] to use systems of asymmetric superconducting loop with or without Josephson junctions as a quantum detector of noise in wide frequency band. The expected dependence of the critical amplitude $I_{0,c}$ of the current or noise, the maximum amplitude $V_{max}$ of $V(\Phi/\Phi_0)$ and $I_{max}$ on temperature, $I_{0,c}(T), V_{max}(T), I_{max}(T)$, allows to calibrate a system of asymmetric superconducting loops as the noise detector. In order to make this calibration the dependencies of the amplitude $V_A$ of the quantum oscillations on the amplitude $I_0$ of the external ac current and the temperature dependence $I_{0,c}(T), V_{max}(T), I_{max}(T)$ should be measured below $T_c$, where the critical amplitude $I_{0,c}(T)$ exceeds noise amplitude. The extrapolation of the measured dependencies $V_A(I_0), I_{0,c}(T), V_{max}(T), I_{max}(T)$ to the region near superconducting transition allows to restore an amplitude profile of the noise pulses from a measured temperature dependence of an amplitude $V_A(T)$ of the quantum oscillations $V(\Phi/\Phi_0)$ induced by this noise.

An asymmetric superconducting loop should have maximum sensitivity as noise detector when a value of the critical current $I_c$ exceeds only slightly a value of the persistent current $I_p$. A low noise current can induce the quantum oscillations of the dc voltage in this case. This condition may be enough easy provided in superconducting loop with Josephson junctions, analogous to SQUID structures[5]. In the present work asymmetric aluminum loops without Josephson junctions is investigated. We did not crave for high sensitivity in this work. It is first systematic investigation of the quantum oscillations of the dc voltage induced by an external ac current at different temperature allowing to make a calibration of the quantum noise detector based on asymmetric superconducting loops (ASL). Analogous to SQUID structures in magnetic field,[5] ASL structures are able to rectify applied ac current or both external and internal noise without bias current when narrow part of the ASL is in the resistive state induced by the magnetic flux quantization persistent current. Single loop and systems connected in series 18 and 20 ASL have been investigated.

## 2. FABRICATION

We use aluminum loops since a large value of superconducting coherence length ξ of aluminum allows to obtain enough high relation of the persistent current to the critical current $I_p/I_c \approx \xi(T)/r$ [8] at a radius r of loop which can be enough easy made by the methods of current technology. Investigated test structures (without Josephson junctions) consisted of asymmetric aluminum loops (45-50 nm thick, thermally evaporated on oxidized Si substrates) with semi-loop width of 200 and 400 nm for narrow and wide parts, respectively. 4 μm diameter single ASL and 20 ASL structure were fabricated by e-beam lithography and lift-off process. Figure 1 and 2 shows SEM images of a single asymmetric superconducting loop and 20 loops structure, respectively. For these structures, the sheet resistance was 0.23 Ω/□ at 4.2 K, the resistance ratio R(300 K)/R(4.2 K)=2.7, and superconducting transition temperature was 1.24-1.27 K. Estimated coherence length ξ(T=0 K) was 170 nm and penetration depth λ(T=0 K) was 80 nm.

## 3. EXPERIMENT

Measurements were carried out by applying 5 or 40 kHz sinusoidal bias current to current leads (fig. 1), whereas, rectified dc signal was measured in a frequency band from 0 to 30 Hz by a home made preamplifier (followed by low-noise preamplifier SR560) at potential leads. Noise level of the amplification system was 20 nVpp for $f_b$=0 to1 Hz. It should be noted that rectification effects do not depend on frequency of the bias current at least up to 1 MHz.[7] Magnetic field direction was perpendicular to the loop's plane. Magnetic field time scanning was slow enough (~ 0.1-1 Hz) so that upper frequency of signal spectrum, which resulted from magnetic field changes, did not exceed 30 Hz. All signals corresponding to rectified voltage, current, temperature and magnetic field were digitized by a multichannel 16-bit analog-to-digital converter card.

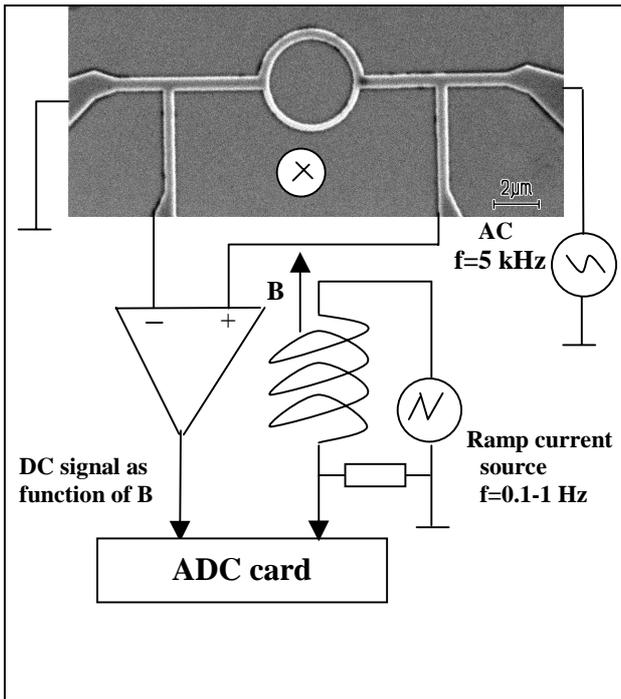

**Figure 1:** An SEM image of a single asymmetric Al superconducting loop and schematics of rectified voltage oscillation measurements as a function of magnetic field.

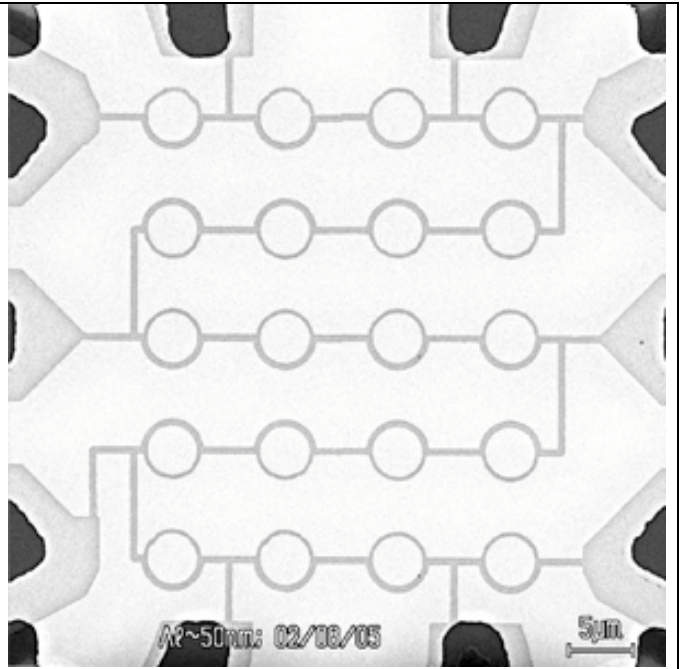

**Figure 2:** An SEM image of a superconducting structure with 20 asymmetric aluminium loops.

To make calibration of the QDN, ac bias current with different amplitude was applied to substitute noise with different amplitude level. Fig.3 shows typical oscillations of the rectified voltage as a function of magnetic field at different ac bias currents for the ASL (T = 1.214 K=0.975$T_c$). The oscillations are periodic with period equals to the superconducting flux quantum $\Phi_o=h/2e$, the positive and negative maximum being observed at $\Phi \approx \Phi_o(n \pm 1/4)$, (n=0,±1,…) showing points of magnetic field with maximum rectification level.

Figure 4 shows that the amplitude of the rectified oscillations in magnetic field for 18 ASLs is approximately 18 times higher compared to that for a single ASL fabricated in the same technological process. This demonstrates a fact that N series connected ASLs synchronously rectify alternating current which results in adding of individual rectified voltages and improvement of rectification efficiency N times.

It can be proposed two interpretations of the observed rectification of the ac current. The first interpretation is based on the likeness of the quantum oscillations of the rectified voltage $V(\Phi/\Phi_0)$, Fig.3,4 and the persistent current $I_p(\Phi/\Phi_0)$[8]. The dc potential difference on a loop segment $V(\Phi/\Phi_0) = R_{asym}I_p(\Phi/\Phi_0)$ can be non-zero if the resistance difference $R_{asym} \neq 0$ and the persistent current $I_p \neq 0$ are not zero. Below superconducting transition, at $T < T_c$, $I_p \neq 0$ but the resistance, and consequently the resistance difference equals zero $R_{asym} = 0$, whereas above $T_c$ $R_{asym} \neq 0$ but $I_p = 0$. Therefore the quantum oscillations $V(\Phi/\Phi_0)$ can be observed only in the fluctuation region near $T_c$ without an external current. The external bias current lowers temperature of superconducting transition. Therefore the quantum oscillations $V(\Phi/\Phi_0)$ induced by the external as current at $T < T_c$ may be interpreted as result of shift $-\Delta T_c(I_0)$ of the critical temperature $T_c$. The quantum oscillations $V(\Phi/\Phi_0)$ is not observed at a low amplitude $I_0$ when $\Delta T_c(I_0) < T_c - T$ and consequently $R_{asym} = 0$. The maximum amplitude, Fig.5, of the $V(\Phi/\Phi_0)$ is observed at $\Delta T_c(I_0) = T_c - T$. The amplitude $V_A$ decreases with further increase of the ac current amplitude $I_0$, Fig.3-5, since $I_p = 0$ above superconducting transition $T > T_c - \Delta T_c(I_{ext})$. The quantum oscillations are observed up to high value current amplitude $I_0 \gg I_{0,c}$ since $|I_{ext}|$ changes in time between 0 and $I_0$ and consequently $\Delta T_c(I_{ext})$ between 0 and $\Delta T_c(I_0)$.

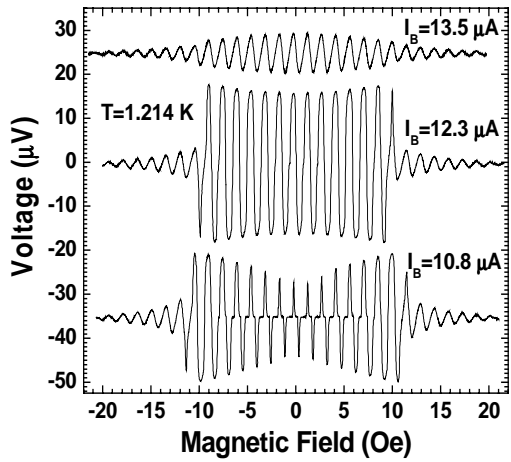
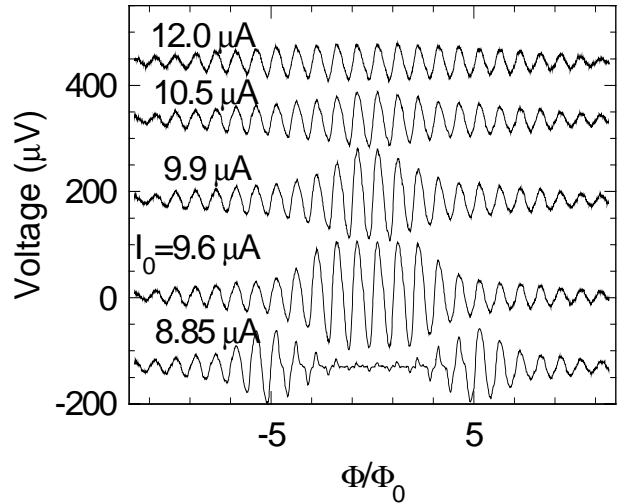

**Figure 3:** The quantum oscillations of the rectified dc voltage for the asymmetric superconducting loop in magnetic field for different ac bias currents (f=5 kHz and amplitudes $I_0$=10.8; 12.3; and 13.5 μA) at T=1.214 K=0.975$T_c$. Except for $I_0 = 12.3$ μA, all curves are vertically shifted.

**Figure 4:** The quantum oscillations of the rectified dc voltage on 18 loops in magnetic field for different ac bias currents (f=40 kHz and amplitudes $I_0$=8.85; 9.6; 9.9; 10.5; and 12 μA) at T=1.244 K=0.98$T_c$. Except for $I_0 = 9.6$ μA, all curves are vertically shifted.

The second interpretation, proposed in[7], explains the $V(\Phi/\Phi_0)$ as a consequence of the asymmetry of the current-voltage curves. Our measurements show that the current-voltage curves of asymmetric aluminum loops are indeed asymmetric when a magnetic flux inside the loop is not divisible to the flux quantum $\Phi \neq n\Phi_0$. The critical currents measured in opposite directions have different value $|I_{ext}|_{c+} \neq |I_{ext}|_{c-}$ and the sign and value of anisotropy $|I_{ext}|_{c+} - |I_{ext}|_{c-}$ changes periodically in magnetic field. Therefore the time average value of the voltage $V_{dc} = \Theta^{-1}\int_\Theta dt V(I_{ext}(t))$ is not zero and its sign and value are periodical function of magnetic flux $V(\Phi/\Phi_0)$. The quantum oscillations $V(\Phi/\Phi_0)$ appears when the ac current amplitude exceeds the minimum value of the critical current, $I_{0,c} = |I_{ext}|_{c+}(\Phi/\Phi_0)$ or $I_{0,c} = |I_{ext}|_{c-}(\Phi/\Phi_0)$. The rectified voltage $V(\Phi/\Phi_0)$ has a maximum value $V_{max}$, Fig.5, when the ac current amplitude is higher than the critical current in a direction $I_0 > \min(|I_{ext}|_{c+}, |I_{ext}|_{c-})$ and lower in the opposite one $I_0 < \max(|I_{ext}|_{c+}, |I_{ext}|_{c-})$, i.e. when the voltage $V(I_{ext}(t))$ has only positive or negative non-zero values. The rectified voltage decreases when non-zero $V(I_{ext}(t))$ values with opposite sign appear at $I_0 > \max(|I_{ext}|_{c+}, |I_{ext}|_{c-})$.

The anisotropy of the current-voltage curves is a consequence of superposition of the external $I_{ext}$ and the persistent $I_p$ current in the narrow $l_n$ and wide $l_w$ loop segments[7]. For example $|I_{ext}|_{c+} = (s_n+s_w)(j_c - I_p/s_n)$ when $I_{ext}$ and $I_p$ have the same direction in the narrow segment with section $s_n$ at positive direction of the external measuring current $I_{ext}$. At opposite direction of the $I_{ext}$ and the same direction of the persistent current $I_{ext}$ and $I_p$ are summed in the wide segment with section $s_w$ and $|I_{ext}|_{c-} = (s_n+s_w)(j_c - I_p/s_w) > |I_{ext}|_{c+}$. The critical currents of superconducting loops are periodical function of the magnetic flux $|I_{ext}|_{c+}(\Phi/\Phi_0)$, $|I_{ext}|_{c-}(\Phi/\Phi_0)$ because of the periodical dependence of the persistent current $I_p(\Phi/\Phi_0)$. The value of the persistent current at different magnetic flux and temperature can be found from the measured $|I_{ext}|_{c+}(\Phi/\Phi_0)$, $|I_{ext}|_{c-}(\Phi/\Phi_0)$ dependencies using the relation $|I_{ext}|_{c-}(\Phi/\Phi_0) - |I_{ext}|_{c+}(\Phi/\Phi_0) = I_p(\Phi/\Phi_0)(s_w/s_n - s_n/s_w)$. The found amplitude of the persistent current is lower in some times that the critical current $I_c = (s_n+s_w)j_c$ of the loop. Therefore one may expect that $I_{0,c} \approx \min(|I_{ext}|_{c+}, |I_{ext}|_{c-})$ and $I_{max} \approx \max(|I_{ext}|_{c+}, |I_{ext}|_{c-})$ should be close to the critical current $I_c$ value.

To determine the exact position of the rectified voltage maximum $I_{max}$ (Fig.5) with respect to the critical current value, in Fig.6 we present temperature dependence of both the critical current of the single ASL obtained from current-voltage characteristics and the amplitude of the bias current $I_{max}$ at which the maximum of the rectified oscillations is observed. As clearly seen from Fig.6, the maximum of the oscillations is achieved when the bias current amplitude is slightly higher than the critical current of the ASL, having practically the same temperature dependence. Temperature dependence of the maximum rectified voltage $V_{max}$ is also shown in Fig.6. The maximum voltage increases with temperature decrease in a way similar to the bias current $I_{max}$, having linear dependence on the bias current (not shown hear). To the accuracy of measurement errors, the ratio $V_{max}/I_{max}=1.45$ Ohm is independent of temperature for $T<0.98T_c$ (fig. 7). When the amplitude of the bias current $I_0$ is equal to $I_{max}$, this ratio determines the maximum rectification efficiency, which is equal to $Eff_{Re} = \sqrt{2}\ V_{max}/I_0R_n = \sqrt{2}\ V_{max}/I_{max}R_n \approx 0.37$, where $R_n=5.5$ Ohm is the resistance of the ASL in the normal state. It should be noted that ASL rectifies sinusoidal current only for one half-period, half-period rectification being characterized by the theoretically limited rectification efficiency of 0.5.

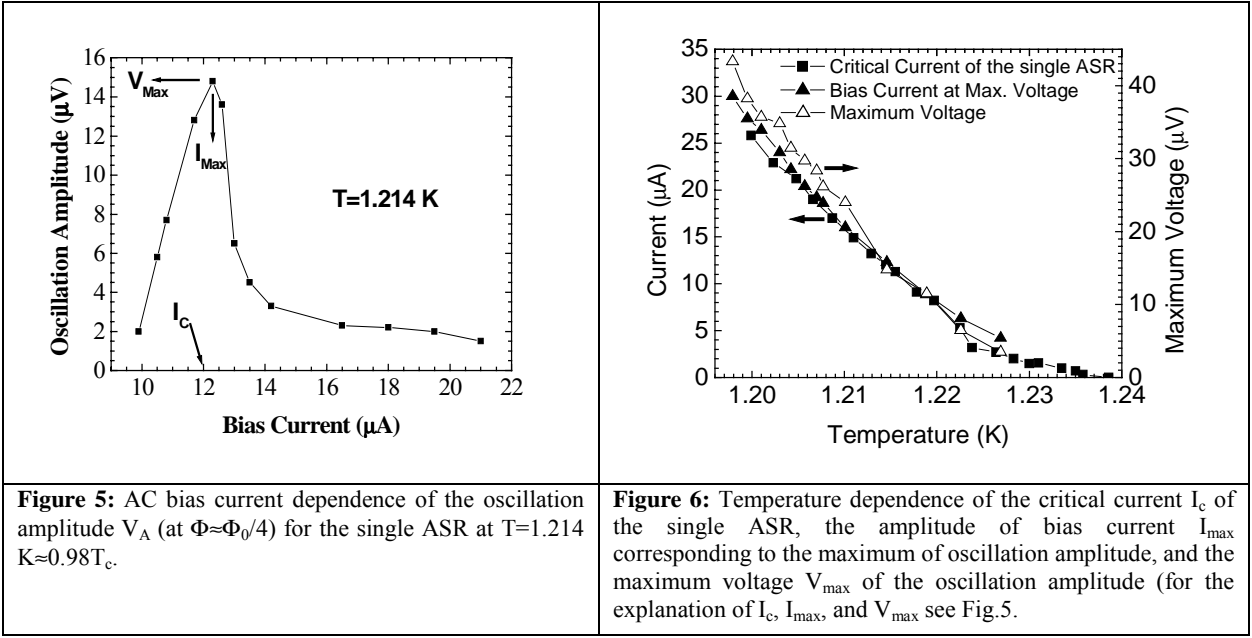

**Figure 5:** AC bias current dependence of the oscillation amplitude $V_A$ (at $\Phi \approx \Phi_0/4$) for the single ASR at $T=1.214$ K$\approx 0.98T_c$.

**Figure 6:** Temperature dependence of the critical current $I_c$ of the single ASR, the amplitude of bias current $I_{max}$ corresponding to the maximum of oscillation amplitude, and the maximum voltage $V_{max}$ of the oscillation amplitude (for the explanation of $I_c$, $I_{max}$, and $V_{max}$ see Fig.5.

Fig. 8 shows the temperature dependence of the maximum rectification efficiency for the single ASL. The high rectification efficiency at low temperatures is determined by the hysteretic current-voltage characteristics (fig. 9). As temperatures approaching $T_c$, the current-voltage characteristics become nonhysteretic resulting in decrease of the maximum rectification efficiency (fig. 10). The rectification efficiency of system 18 and 20 loops $Eff_{Re} = \sqrt{2}\ V_{max}/I_{max}R_n$ is close to the one of the single ASL, fig.10 at low temperature although the resistance $R_n$ = 92 Ohm of system 18 loops and $R_n$ = 58 Ohm of 20 loops is higher than the resistance $R_n$=5.5 Ohm of single loop. This means that the relation $V_{max}/I_{max} \approx 20$ Ohm for 18 loops is higher then for $V_{max}/I_{max}$=1.45 Ohm for the single ASL.

Our preamplifier noise level allows reliable measurement of 30 nV voltage signal and therefore we can see a noise with current amplitude down to 1.5 nA using the system of 18 loops. But it could be possible only in the temperature region near superconducting transition $T_c$ where the critical current is lower than 1.5 nA. But the rectification efficiency of asymmetric aluminum loops decreases strongly in this region, fig.10. There is important to note that the current-voltage characteristics are nonhysteretic near superconducting transition, at low critical current. The rectification efficiency of a structure with the nonhysteretic current-voltage characteristics can not exceed the value $Eff_{Re} \approx 0.5[1 - (|I_{ext}|_{c,min}/|I_{ext}|_{c,max})^2]$. The relation of the critical currents according to our measurements $|I_{ext}|_{c,min}/|I_{ext}|_{c,max} \approx 0.82 - 0.70$. Therefore the rectification efficiency should be lower than $Eff_{Re} \approx 0.16 - 0.25$. The measured rectification efficiency is higher this value at low temperature, where the current-voltage characteristics are hysteretic, Fig.9, and is lower it near superconducting transition, Fig.10, where the current-voltage characteristics are nonhysteretic and more smooth.

Our measurements have shown that the temperature dependencies of the critical amplitude $I_{0,c}(T)$ and the amplitude of the bias current $I_{max}(T)$ at which the maximum of the rectified oscillations is observed are close to the one of the critical current $I_c(T)$ of the loops, fig.6. Using this similarity $I_{0,c}(T) \approx I_c(T)$, $I_{max}(T) \approx I_c(T)$, and measured temperature dependencies of the critical current $I_c(T)= I_c(0)\times(1 - T/T_c)^{3/2} = 1.4$ mA$\times(1 - T/T_c)^{3/2}$ of system of 18 ASL, $I_c$=3.0 mA$\times(1 - T/T_c)^{3/2}$ of system of 20 ASL and $I_c$=4.4 mA$\times(1 - T/T_c)^{3/2}$ it is possible to estimate the minimum amplitude of noise which can induced the quantum oscillations of the dc voltage $V(\Phi/\Phi_0)$ at different temperatures $T/T_c$.

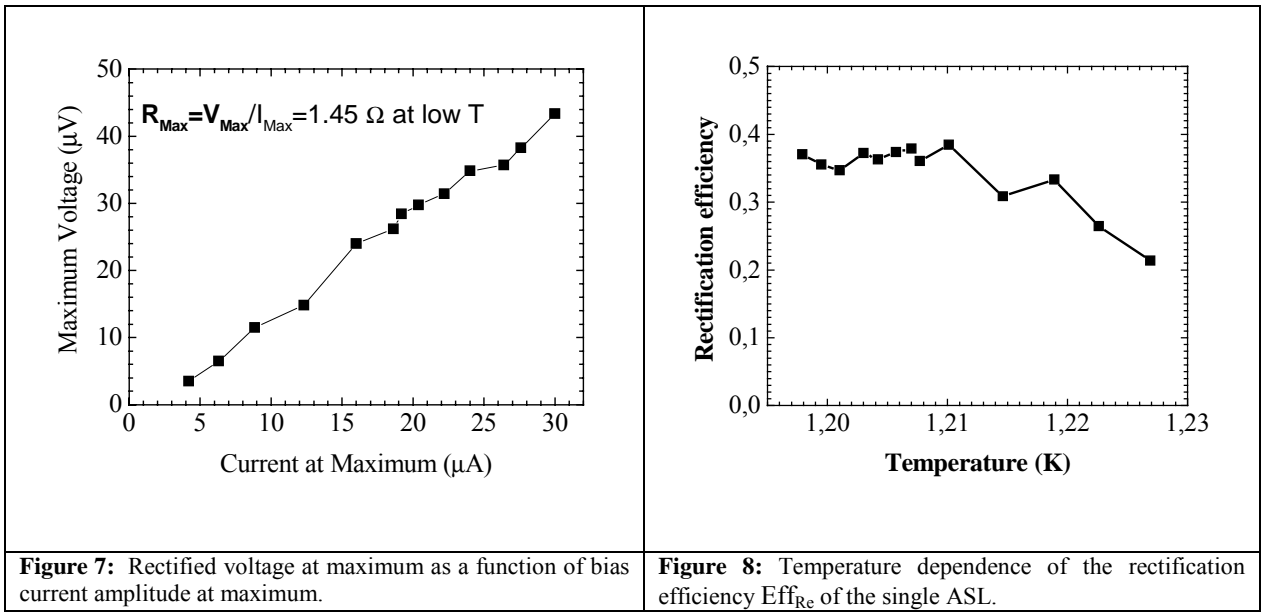

**Figure 7:** Rectified voltage at maximum as a function of bias current amplitude at maximum.

**Figure 8:** Temperature dependence of the rectification efficiency $Eff_{Re}$ of the single ASL.

Since the bias current (or noise which is equivalent to bias current) only shifts the QDN total current to near critical value $I_{max}$, where it has the maximum rectification efficiency, detection of noise at different temperatures near $T_c$ allows amplitude profiling of the noise pulses in accordance with $I_c(T)$ temperature dependence. Therefore, the noise with the lower level, both equilibrium and non-equilibrium, will be detected by the QDN without biasing at nearer to $T_c$ temperatures. Let us estimate where the QDN can detected the equilibrium noise. The current level of the equilibrium noise can be determined as root mean square of the power $R<I_{Ny}^2> =k_BT\Delta\omega$ of the Nyquist noise which is uniformly distributed among the frequency range from 0 to the quantum limit $k_BT/\hbar = 1.3\ 10^{11}$ Hz at $T \approx 1$ K, i.e. the full power of the Nyquist noise $W_{Ny} \approx (k_BT)^2/\hbar$ and the maximum value of the equilibrium current $I_{Ny,max} = <I_{Ny}^2>^{1/2} \approx k_BT/(\hbar R)^{1/2}$ is equal approximately ~0.4 µA at T ≈ 1 K and R ≈ 10 Ohm, for example. The time of current relaxation in the ASL can be estimated by the value $R/L \approx 10^{-12}$ s. The relaxation time in equilibrium superconducting state equals $\tau_0 = \pi\hbar/8k_B(T_c-T)$ One may expect that the noise in the frequency band from 0 to $8k_B(T_c-T)/\pi\hbar$ can induce the dc voltage $V(\Phi/\Phi_0)$. The average amplitude of the noise current in this band $\langle I_{Ny}^2\rangle^{1/2} \approx (8/\pi)^{1/2}(1 - T/T_c)^{1/2} I_{Ny,max} \approx 0.6$ µA $(1 - T/T_c)^{1/2}$ at T ≈ 1 K and R ≈ 10 Ohm. The critical current $I_c(T) = I_c(0) \times (1 - T/T_c)^{3/2}$ becomes lower this value at $1 - T/T_c = (8/\pi)^{1/2} I_{Ny,max}/I_c(0)$ = 0.6 µA/$I_c(0)$, i.e. at $1 - T/T_c \approx 0.0005$ for the system of 18 loops with $I_c(0) = 1.4$ mA, at $1 - T/T_c \approx 0.0002$ for the system of 20 loops with $I_c(0) = 3.0$ mA, at $1 - T/T_c \approx 0.00014$ for the single ASL with $I_c(0) = 4.4$ mA. This values are much smaller than the width of the resistive transition, for example $\Delta T/T_c \approx 0.02$ of the system of 18 ASL, and the width of the temperature region where the value $Eff_{Re}$ of the rectification efficiency changes, fig.10.

Thus, we can expect that the QDN based on ASR investigated in our work can not detect equilibrium noise. There are two principal defects of system of asymmetric aluminum loops used in our work as QDN: 1) the value of the critical current is too high and 2) the rectification efficiency changes strongly near superconducting transition because of change of the current-voltage curves. These defect can be removed by using asymmetric superconductin loops with Josephson junctions which diminish the critical current, i.e. structures like rf-SQUID or asymmetric dc-SQUID.

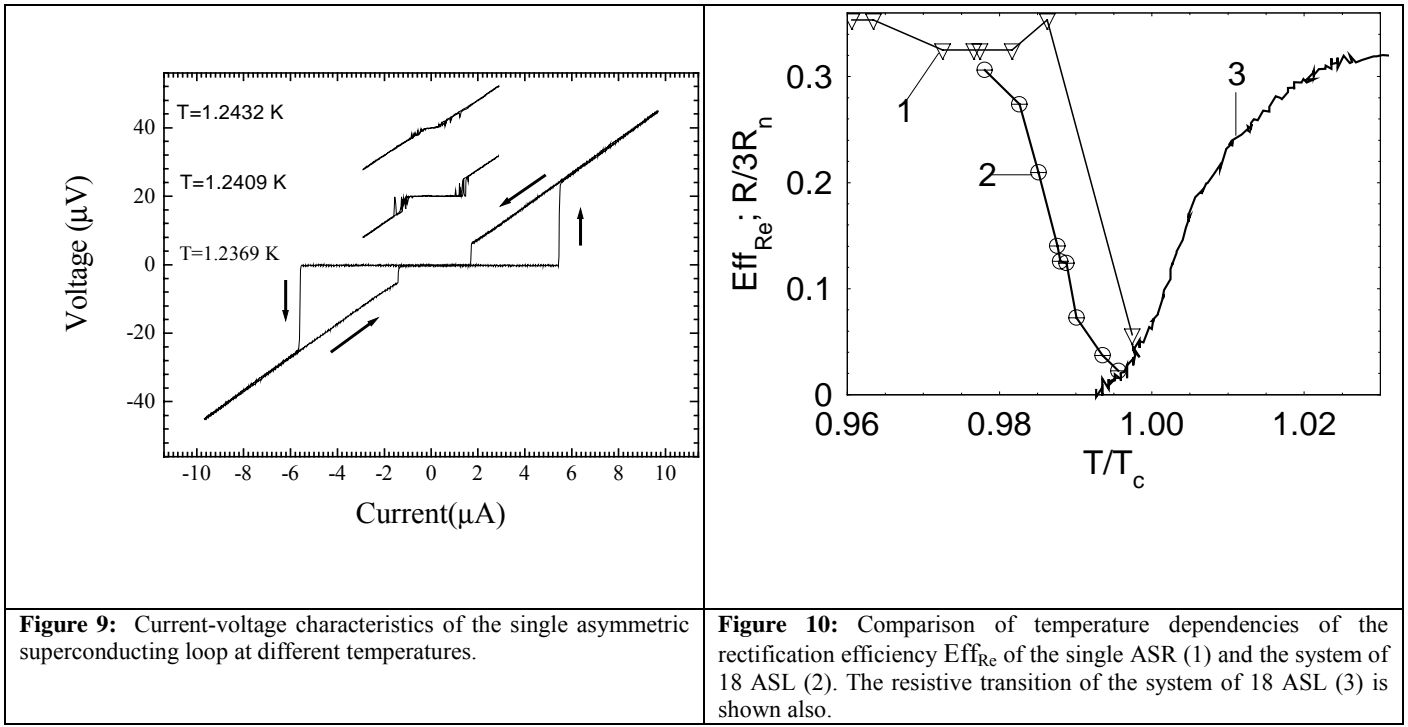

**Figure 9:** Current-voltage characteristics of the single asymmetric superconducting loop at different temperatures.

**Figure 10:** Comparison of temperature dependencies of the rectification efficiency $Eff_{Re}$ of the single ASR (1) and the system of 18 ASL (2). The resistive transition of the system of 18 ASL (3) is shown also.

## CONCLUSION

Calibration of the noise detector and optimization of the sensitivity has been carried out. Sensitivity improvement has been achieved in multiple series connected loop structures conforming synchronous rectification of the alternating signal which results in adding of rectified voltages from individual loops. Rectification of noise at different temperatures near superconducting critical temperature allows amplitude profiling of the noise pulses.

We can conclude that although a system of asymmetric aluminum loops can be used as quantum detector of noise it is not best system. The value is determined by the relation of the persistent current to the critical current. Therefore the maximum rectification efficiency could be increased using weak links or Josephson junctions which diminish the critical current (rf-SQUID or asymmetric dc-SQUID structure). One may expect that a system of asymmetric loops with

Josephson junctions could be the quantum noise detector of wide frequency band from 0 to $8k_B(T_c-T)/\pi\hbar$ with highest sensitivity.

This quantum noise detector can be used, for example, for the control of an external noise at the investigation of the possibility macroscopic quantum tunneling and macroscopic quantum superposition. The statements on experimental evidence of macroscopic quantum tunneling in[17-19] and other works raise doubts because of the absence of the careful control of non-equilibrium noise[20]. The investigation of the possibility of macroscopic quantum tunneling is very important. This fundamental problem is connected directly with the problem of possibility of a real quantum computer.

**Acknowledgments**
This work was financially supported by ITCS department of RAS in the Program "Technology Basis of New Computing Methods", by Russian Foundation of Basic Research (Grant 04-02-17068) and by the Presidium of Russian Academy of Sciences in the Program "Low-Dimensional Quantum Structures".